\documentstyle[twocolumn,aps]{revtex}
\begin{document}
\bibliographystyle{unsrt}
\title{Ordered and Disordered Defect Chaos}
\author{Glen D. Granzow and Hermann Riecke}
\address{Department of Engineering Sciences and Applied Mathematics\\
Northwestern University, Evanston, IL 60208, USA}
\maketitle

\begin{abstract}
Defect-chaos is studied numerically in coupled Ginzburg-Landau equations 
 for parametrically driven waves.  The motion of the defects is traced in detail
yielding their life-times, annihilation partners, and distances traveled. 
 In a regime in which in the one-dimensional
case the chaotic dynamics is due to double phase slips, the two-dimensional
system exhibits a strongly ordered stripe pattern. When the parity-breaking
instability to traveling waves is approached this order vanishes and the correlation
function decays rapidly. In the ordered regime the defects have a typical life-time, 
whereas in the disordered regime the life-time distribution is exponential. 
The probability of large defect loops is substantially larger in the disordered regime.
 
\end{abstract}

\centerline{April 19, 1997}
\ \\
Paper for the proceedings of the V$^{th}$ Bar-Ilan Conference {\it Physics of 
Complex Systems}.



\section{Introduction} 
\label{sec:intro}
 
Spatio-temporal chaos has been studied in quite some detail experimentally 
and theoretically. One of the difficulties in studies of such complex states
is the identification
of quantities that allow a suitable characterization of the complex
dynamics and of transitions that might occur between different types
of dynamics. A large number of theoretical investigations has focussed on the
complex Ginzburg-Landau equation, which describes the dynamics of
weakly nonlinear waves. There, spatio-temporal chaos arises over a wide
range of parameters. In one dimension two regimes have been identified,
one (amplitude chaos) in which the dynamics is dominated by phase slips, 
i.e. events during which the
amplitude of the wave goes to zero and 
a wavelength is inserted or eliminated, and a regime in which
(essentially) no phase slips occur (phase chaos) \cite{ShPu92,EgGr95}. 
In two dimensions the role of the phase slips
is played by the creation and annihilation of defects in the wave pattern,
which in this system have the form of spirals \cite{CoLe88}. 
A number of different regimes of such spatio-temporal chaos 
have been identified \cite{ArKr94,ChMa96}.

Since the spirals are topologically
stable and easily identified it is natural to attempt a characterization of the
chaotic state in terms of the dynamics and statistics of these defects.
It has been found that the probability distribution function for the number of
defect pairs is consistent with a model in which the defects are
created randomly in pairs and annihilated by each other at a rate that
is proportional to their density, i.e. they appear to be statistically independent
of each other \cite{GiLe90}. This was also found in 
experiments on electro-convection in nematic liquid crystals \cite{ReRa89}. 
Since the defects are always created in pairs of opposite `charge' 
the total charge is conserved and the charge density is expected to 
evolve on large space- and time-scales. This was investigated in \cite{RoBo96}.
There, it was also remarked that the defects are typically not bound in pairs,
i.e. they are usually annihilated by a defect that is not the defect they
were created with. No detailed analysis of this question has been performed,
however.

In this paper two-dimensional spatio-temporal chaos is investigated in 
a model for parametrically excited waves. The work is motivated by previous results
in one dimension, which identified a regime in which the dynamics is 
due to double phase slips rather than single phase slips \cite{GrRi96,GrRi97}.
Double phase slips consist of a sequential pair of phase slips in which
 the second phase slip quickly follows and negates the first. Across a double phase
slip the total phase of the system is conserved. Thus, the waves have a well-defined
effective wave number and a large-scale description of the
chaotic state with an effective phase diffusion equation is possible \cite{GrRi96,GrRi97}. 
In two dimensions the double phase slips are expected to correspond to 
`fluctuating bound defect pairs',
i.e. to pairs of defects that are created together and are annihilated by each other
after a typical time during which the defects remain strongly correlated. Preliminary
computations have indicated such a behavior \cite{GrRi97}. An interesting question
concerns the possibility of an `unbinding' of these defect pairs and as a consequence
of a transition from a more ordered to a less ordered chaotic state. 
Here we present computations which show
a drastic decrease in the order of the chaotic state connected with the disappearance
of the typical lifetime of the defects and other signatures in the defect statistics.

\section{The Model Equations}

In this paper we study the dynamics of parametrically excited waves within the
framework of coupled Ginzburg-Landau equations. These describe the temporal evolution
of the complex amplitudes $A$ and $B$ of oppositely traveling waves 
that are coupled due to a 
parametric forcing at a frequency $\omega_e$, which is close to twice the frequency
 of the traveling waves. The traveling waves are assumed to arise
in a supercritical Hopf bifurcation. In terms of 
the amplitudes $A$ and $B$, physical quantities 
$u$ like the vertical fluid velocity in the mid-plane of a convection 
system are given by 
\begin{eqnarray}
u({\bf r},t)=\epsilon A ({\bf R},T)e^{i({\bf q}_c \cdot {\bf r}-\frac{\omega_e}{2}t)} 
      +\epsilon B ({\bf R},T)e^{i({\bf q}_c \cdot {\bf r}+\frac{\omega_e}{2}t)}\nonumber \\
      +c.c.+o(\epsilon).                                 \label{e:defu}
\end{eqnarray}
The amplitudes $A$ and $B$ vary on the slow time and space scales, 
$T=\epsilon^2t$, ${\bf R}\equiv (X,Y)=\epsilon {\bf r}$ with ${\bf r}=(x,y)$ and
$\epsilon \ll 1$.

The ansatz (\ref{e:defu}) introduces an anisotropy into 
the system since the slowly varying coordinate ${\bf R}$ 
does not capture large variations in the orientation of the waves. 
This leads to difficulties in the 
description of isotropic physical systems. For their description one is therefore
either forced to restrict oneself to situations in which the orientation of the
wavenumber varies only slightly or one has to allow the amplitudes to vary on the
fast scale ${\bf r}$. In the former case a systematic derivation leads to coupled
Newell-Whitehead-Segel equations for the amplitudes (e.g. \cite{CrHo93}). 
In the latter case one obtains
a complex order-parameter equation in which the nonlinearities contain complicated
integral terms (e.g. \cite{NeFr93}). Usually, these are then approximated by 
local terms (e.g. \cite{NeFr93,ZhVi95}).

If the physical system has an axial anisotropy no such difficulties arise since
the waves travel essentially along the preferred direction as is, for instance, the
case in electro-convection in nematic liquid crystals \cite{ReRa89,ReRa88,DeTr96}. 
In that case, if the waves arise in a supercritical bifurcation, they can be described 
by coupled Ginzburg-Landau equations of the form \cite{RiCr88,Wa88}
\begin{eqnarray}
\partial_TA+s\partial_XA &=& d\nabla^2A+aA+bB\nonumber \\
 +cA(|A|^2+|B|^2)+gA|B|^2,                               \label{e:cglA} \\
\partial_TB-s\partial_XB &=& d^*\nabla^2B+a^*B+bA\nonumber \\
 +c^*B(|A|^2+|B|^2)+g^*B|A|^2.                           \label{e:cglB}
\end{eqnarray}
The coefficients in (\ref{e:cglA},\ref{e:cglB}) are 
complex except for $s$ and $b$, which are real.
The real part $a_r$ of the coefficient $a$ gives the linear 
damping of the traveling waves in the absence of the periodic forcing and is
proportional to the distance from the Hopf bifurcation. 
The coefficient of the linear coupling term $b$ gives the amplitude of the 
periodic forcing as can  be seen from the fact that it 
 breaks the continuous time-translation symmetry $t \rightarrow t+\Delta t$, which
implies the transformation $A \rightarrow Ae^{i\Delta t \omega_e/2}$
$B \rightarrow Be^{-i\Delta t \omega_e/2}$.
The imaginary part $a_i$ of the coefficient $a$ gives the 
difference between the frequency of the unforced waves (at wavenumber ${\bf q}_c$)
and half the forcing frequency $\omega_e$.

In addition to the trivial solution $A=B=0$, (\ref{e:cglA},\ref{e:cglB}) possess
three types of simple solutions: $|A|=|B|=const$, $|A| \ne |B|$ (both constant), and
$|A|=|B|$ (both time-periodic) \cite{RiCr88}. 
We are in particular interested in the first type,
which corresponds in the physical system to standing waves that are phase-locked to the
parametric forcing, i.e. they are excited by the forcing. With increasing $a_r$ they
become unstable to solutions of the second type, which correspond to traveling waves 
as they exist also in the absence of any periodic forcing. Solutions of the third
type correspond to standing waves that are not phase-locked to the forcing. For 
the case discussed here, $g_r<0$, they are unstable to the traveling waves.  

\begin{figure}[p] 
\begin{picture}(420,270)(0,0)
\put(-50,-50) {\includegraphics{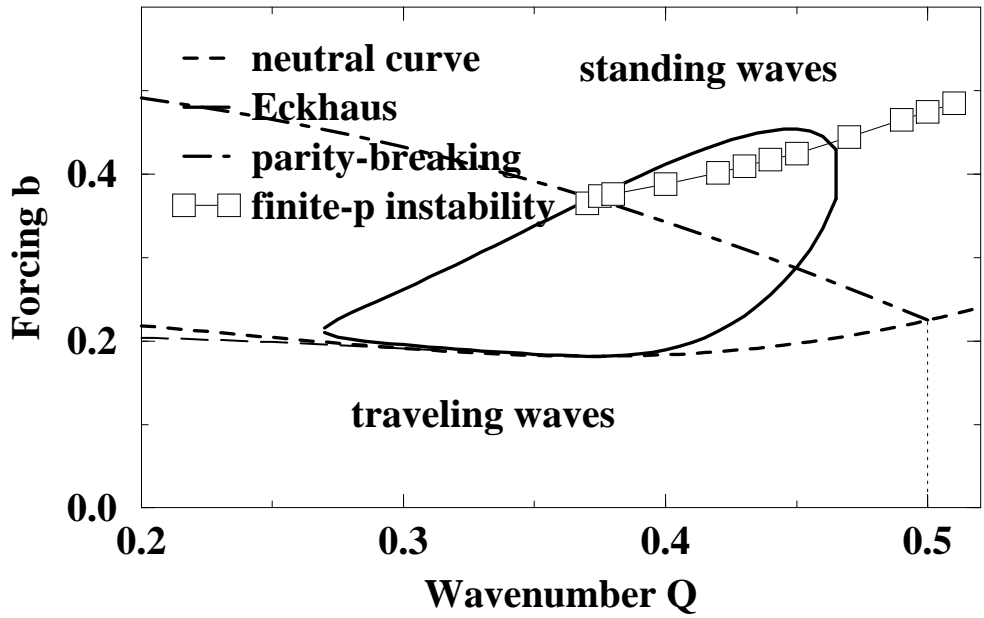}}
\end{picture}
\caption{Linear stability diagram for  
$a=0.25$, $c=-1+4i$, $d=1+0.5i$, $s=0.2$, $g=-1-12i$. For details see text.
\protect{\label{f:stabi}}
}
\end{figure}

The stability of the phase-locked standing waves with respect to long-wave modulations
(Eckhaus instability) can be described using a phase-diffusion equation.
The stability limits are then given 
by a sign-change of the diffusion coefficient. For the one-dimensional case 
the diffusion coefficient was determined analytically in \cite{Ri90a}. 
The phase-diffusion equations does not describe the parity-breaking 
instability, at which the standing waves become unstable to traveling waves \cite{RiCr88}.
Fig.\ref{f:stabi} shows the stability limits relevant for the present paper.
The neutral curve, above which the trivial state is unstable to standing waves,
 is given by the dashed line. Their Eckhaus instability is denoted by a solid line
and their parity-breaking instability by a dashed-dotted line. Over some range
of parameters the parity-breaking instability is preempted by a mode that arises first
at finite modulation wavenumber (open squares). It emerges from the parity-breaking
instability. The standing waves are 
stable only inside the region marked by the solid lines and the squares. For completeness
it should be noted that to the left of the minimum of the neutral curve the bifurcation
to standing waves becomes subcritical; the saddle-node line is shown as a thin long-dashed line.
For negative $Q$ a second minimum of the neutral curve exists, but the only stable regime
of the standing waves is as shown in fig.\ref{f:stabi}.
Traveling waves exist below the parity-breaking instability and to the left of 
the neutral curve indicated by the dotted line at $Q=0.5$.

As in the case studied in 
\cite{Ri90a,GrRi96,GrRi97}, in which $a_r<0$ and no traveling waves arise,
 the region of stable standing waves is closed in fig.\ref{f:stabi}
($a_r=0.25$) with the standing waves becoming 
Eckhaus-unstable for large forcing. In one dimension, 
it was found that for larger $b$ the Eckhaus
instability does not lead to a single phase slip, which would change the overall 
wavenumber of the standing waves and would thus take the unstable waves into the
stable band, but instead to double phase slips. They consist of a sequential pair
of phase slips with the second phase slip undoing the action of the first one.
Therefore the total phase is conserved across such a double phase slip 
and the overall wavenumber remains the same. 
As a consequence, persistent dynamics - regular and chaotic - can 
arise. In particular, it was found that the chaotic dynamics can become localized
in space, a phenomenon that can be understood within the framework
of an effective phase diffusion equation \cite{GrRi96,GrRi97}.

In the complex Ginzburg-Landau equation describing unforced traveling waves no regime
with double phase slips seems to exist. With increasing $a_r$ the parametrically excited
standing waves become unstable to traveling waves in a parity-breaking instability
\cite{RiCr88,ReRa88}. The latter are only weakly affected by the forcing for 
sufficiently large $a_r$. One may therefore expect an interesting transition
from chaotic dynamics characterized by double phase slips to one dominated by single
phase slips when $a_r$ is increased towards the parity-breaking instability.

In two dimensions the overall wavenumber is changed by dislocations. A
single phase slip is effected by a pair of dislocations being created and
separating to infinity, i.e. disappearing at opposite walls of the
system. Since the two phase slips in a double phase slip undo each
other one might expect that the double phase slips correspond to a process 
in which two dislocations are created and move apart, but then turn around
and annihilate each other again. Thus, they would appear as if they were
bound to each other. Preliminary calculations have shown this behavior 
\cite{GrRi97}.

\section{Ordered and Disordered Chaotic Waves}

\begin{figure}[p] 
\begin{picture}(420,270)(0,0)
\put(60,0) {\includegraphics{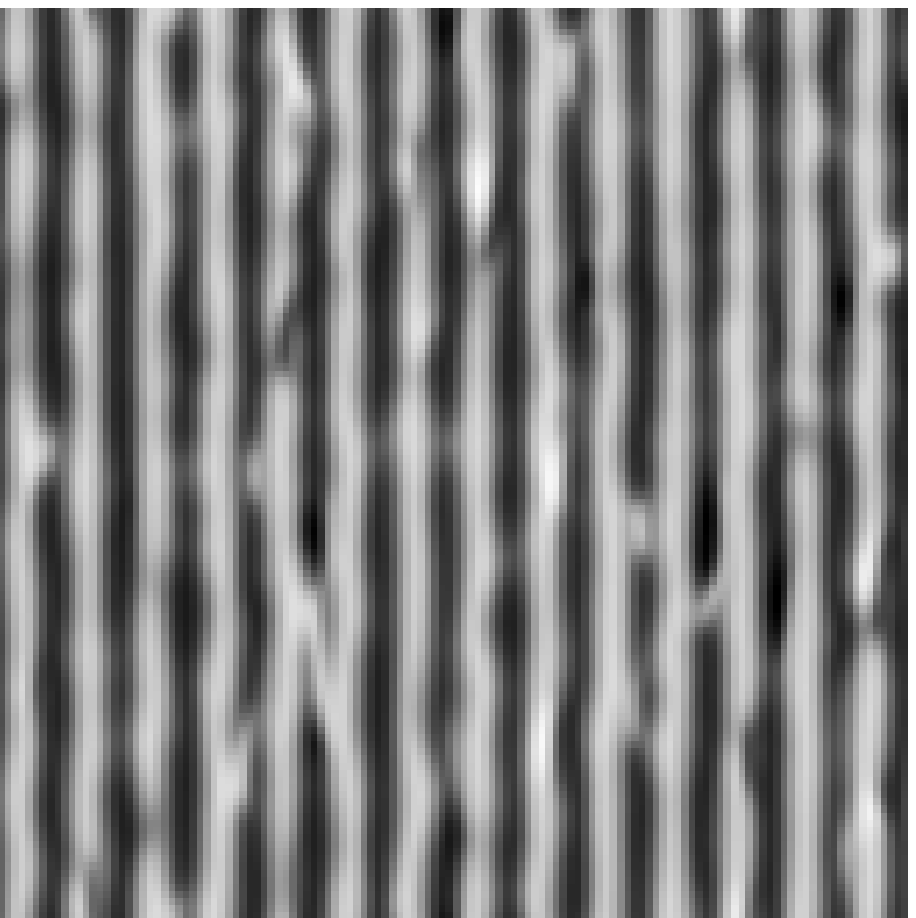}}
\end{picture}
\caption{Snapshot of the real part of the wave pattern for $b=1$ (other parameters as in
fig.\protect{\ref{f:stabi}}).\protect{\label{f:snapb1}}
}
\end{figure}

\begin{figure}[p] 
\begin{picture}(420,270)(0,0)
\put(60,0) {\includegraphics{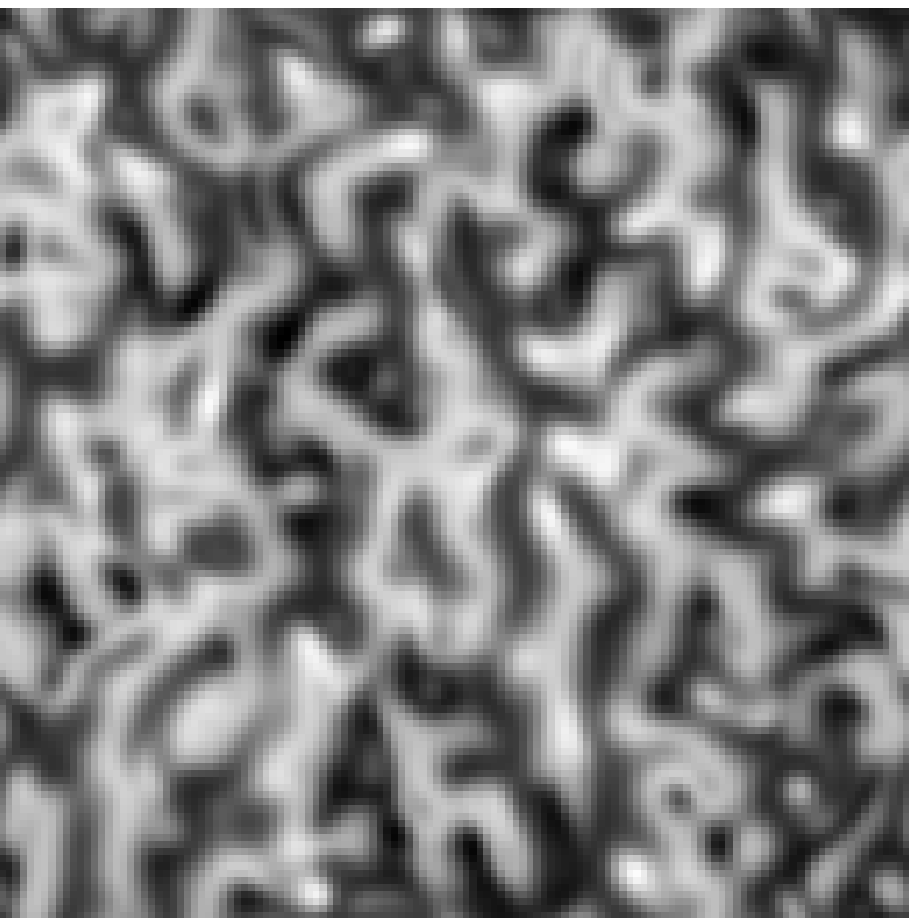}}
\end{picture}
\caption{Snapshot of the real part of the wave pattern for $b=0.5$ (other parameters as in
fig.\protect{\ref{f:stabi}}).\protect{\label{f:snapb05}}
}
\end{figure}

To investigate the dynamics of defects in detail we simulate (\ref{e:cglA},\ref{e:cglB})
numerically using a pseudo-spectral method with a $4^{th}$-order 
Runge-Kutta/integrating factor time-stepping scheme. The time step is $dt=0.25$ and
the highest Fourier mode is $N=64$ for a system length $L=136$.
Periodic boundary conditions are used. The parameters are chosen as
in fig.\ref{f:stabi}. Since $a_r >0$, the standing waves become unstable to traveling
waves when the forcing $b$ is decreased. It is expected that this parity-breaking 
instability will also affect the dynamics of the defects, in particular
since in 1 dimension the traveling waves do not exhibit any double
phase slips. 
Here we present results for $b=1$, i.e. far away from the parity-breaking
instability, and for $b=0.5$, which is much closer to 
that instability.
A typical snapshot of the waves is shown in fig.\ref{f:snapb1} for $b=1$
and in fig.\ref{f:snapb05} for $b=0.5$. 

\begin{figure}[p] 
\begin{picture}(420,270)(0,0)
\put(50,-90) {\includegraphics{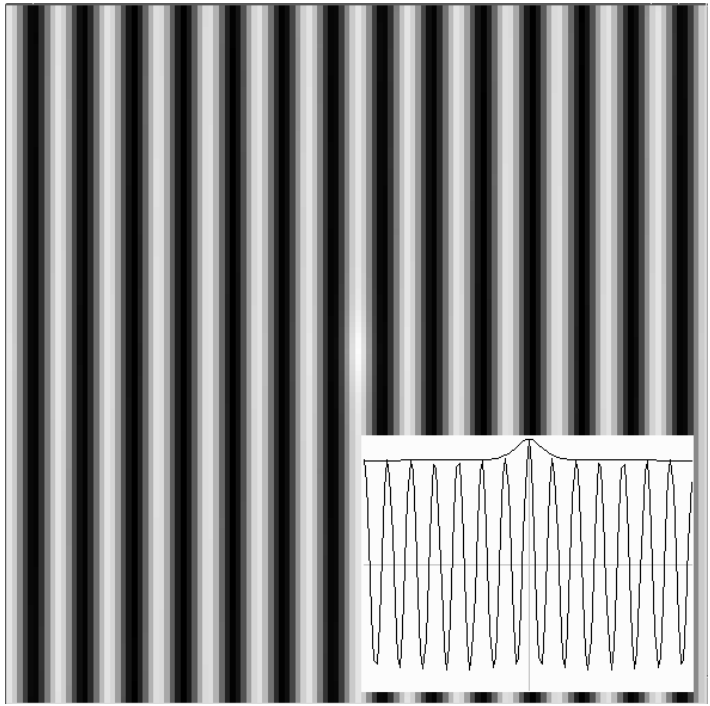}}
\end{picture}
\caption{Time-averaged correlation function of the wave pattern for $b=1$. The inset shows
 one-dimensional cuts in the $X$- and $Y$-direction on a linear scale. The
other parameters are as in
fig.\protect{\ref{f:stabi}}.\protect{\label{f:corb1}}
}
\end{figure}

For $b=1$ the wave pattern consists of
a quite ordered array of stripes with a number of defects. The
correlation function in fig.\ref{f:corb1}, which is an average over $\Delta T=20000$ 
shows that
in this regime the correlation function decays only very little over the whole system. 
In analogy to the one-dimensional chaotic state that is driven by double phase slips,
it is expected that the large-scale dynamics of this stripe pattern can be described by 
a diffusion equation with noise \cite{GrRi97}. 
No true long-range order is expected in such a stripe
pattern. For the system
sizes investigated  so far ($L=68$ and $L=136$), we could not address this question. 
Since the pattern
has a quite well-defined wavenumber the dynamics depends somewhat 
sensitively on the length, which discretizes the possible wavenumbers.
When the effective wavenumber is decreased (by increasing the length of the system)
 the pattern develops into a zig-zag pattern, very similar to what is observed in 
non-chaotic systems. To avoid this complication we have adjusted the
length such that no zig-zags were apparent. 

\begin{figure}[p] 
\begin{picture}(420,270)(0,0)
\put(50,-90) {\includegraphics{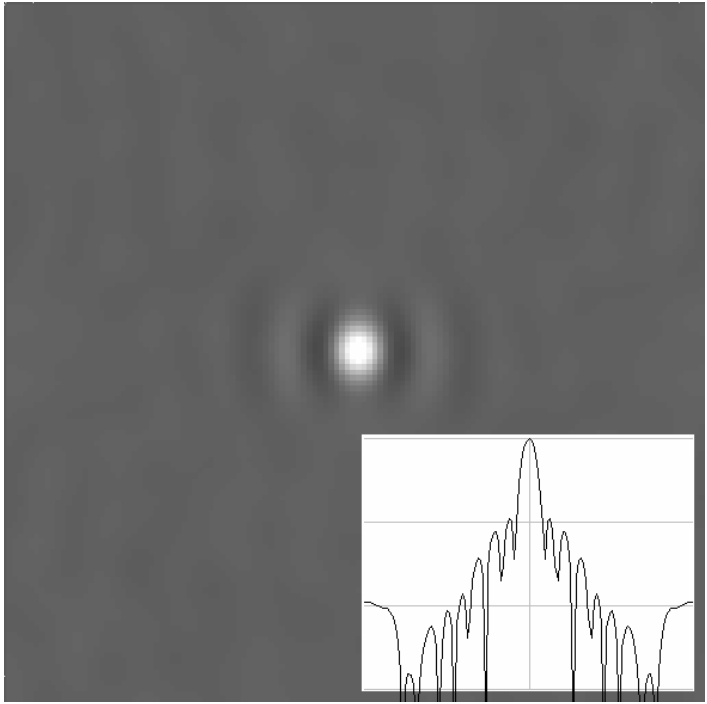}}
\end{picture}
\caption{Time-averaged correlation function of the wave pattern for $b=0.5$.
The inset shows
a one-dimensional cut in the $X$-direction of the absolute value of the correlation
function on a linear-logarithmic scale.
The other parameters are as in
fig.\protect{\ref{f:stabi}}.\protect{\label{f:corb05}}
}
\end{figure}

For $b=0.5$ the wave pattern is quite disordered. The correlation
function in fig.\ref{f:corb05} (averaged over $\Delta T=10000$) 
shows that the pattern becomes
statistically almost isotropic in this case (despite the small
anisotropy due to the group velocity $s=0.2$). The decay of the 
correlation function is clearly exponential.

To characterize the dynamics of the defects they are tracked from 
their creation to their annihilation. In order to address the question
whether there are `bound' defect pairs each defect pair is labeled
and it is noted which defect is annihilated by which other defect. In space-time
the defects trace out loops \cite{GrRi97}. If a defect is annihilated by the same
defect with which it was created such a loop consists only of this
defect pair and is opened when they are created and closed when
they annihilate each other. In general, the loops can consist of any
number of defect pairs.

\begin{figure}[p] 
\begin{picture}(420,270)(0,0)
\put(-50,-50) {\includegraphics{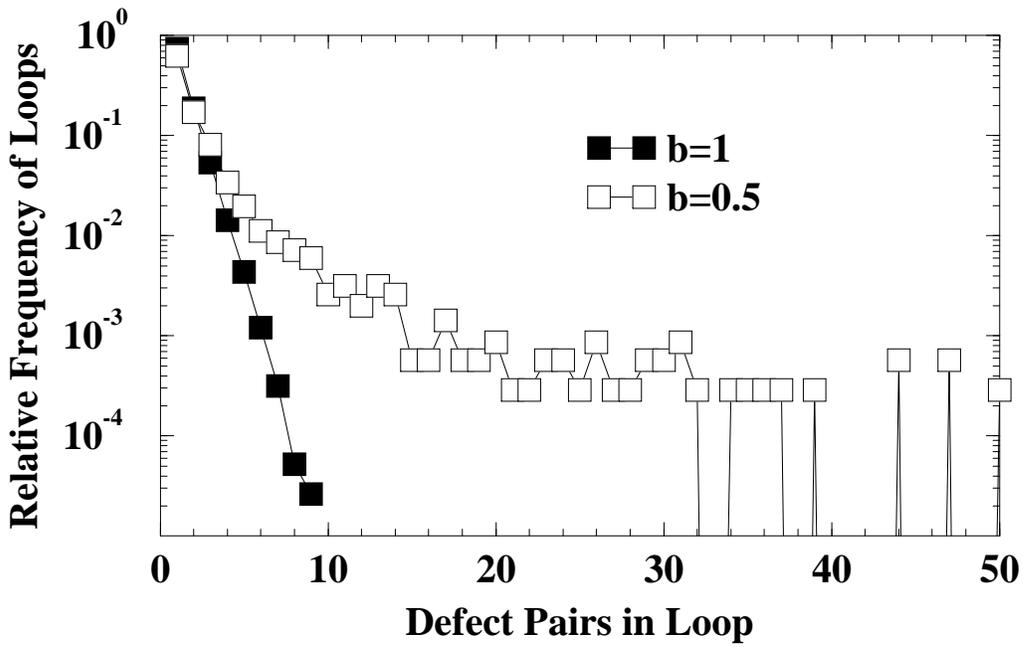}}
\end{picture}
\caption{Relative frequency of defect loops in space-time containing $n$ defect pairs 
for $b=1$ (solid squares) and $b=0.5$ (open squares).
The other parameters are as in
fig.\protect{\ref{f:stabi}}.\protect{\label{f:loop}}
}
\end{figure}

In fig.\ref{f:loop} the probability distribution function for the
number of defect pairs making up a loop is shown for $b=1$ and $b=0.5$.
In both cases most loops contain only a single defect pair. At first
glance this might suggest that the defects are bound to each other in both cases.
This is, however, not the case as the other diagnostics indicate. In fact,
even in the single Ginzburg-Landau equation, in which no double phase slips occur
in 1 dimension and for which no bound defects are expected, 
most loops contain only a single defect pair \cite{GrRiun}. A strong difference between 
the two regimes arises, however, in the probability of multi-loops:  
loops containing many defects are much more likely for $b=0.5$ than for $b=1$.
In fact, from the data available so far 
it is not clear whether the probability
decays exponentially for $b=0.5$.

\begin{figure}[p] 
\begin{picture}(420,270)(0,0)
\put(-50,-50) {\includegraphics{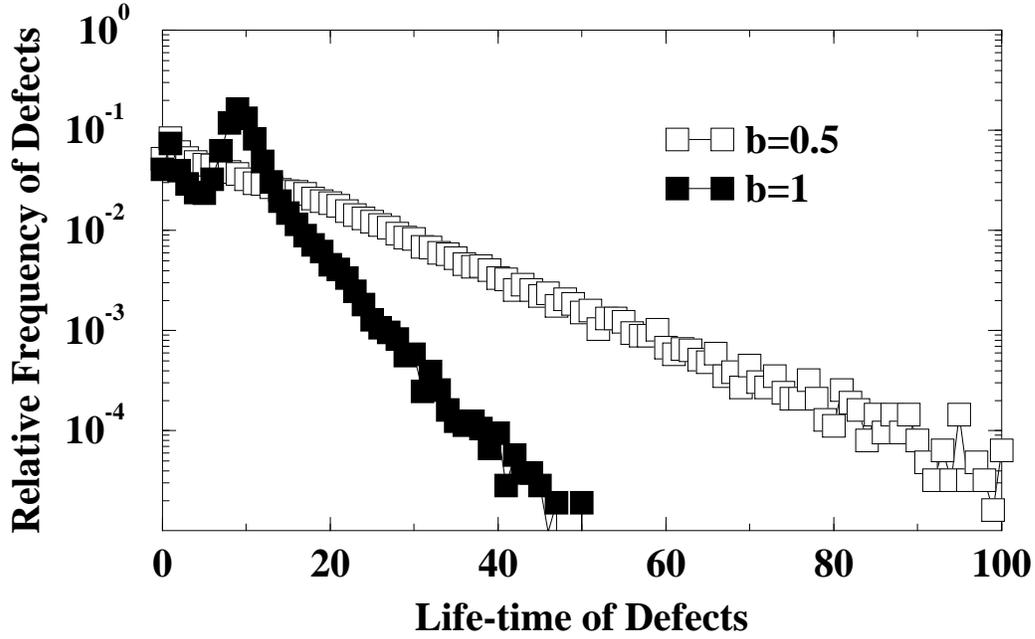}}
\end{picture}
\caption{Relative frequency of defect loops in space-time with life-time $\tau$ 
for $b=1$ (solid squares) and $b=0.5$ (open squares).
The other parameters are as in
fig.\protect{\ref{f:stabi}}.\protect{\label{f:life}}
}
\end{figure}

Another striking difference in the two regimes is noticed in the
distribution of the life-times of each defect. As seen in fig.\ref{f:life}
the distribution has a clear peak for $b=1$ indicating a typical
life-time of the individual defects while for $b=0.5$ the distribution
is exponential over the whole range. The latter indicates that during each time interval
a defect has a certain probability to be annihilated; correlations
between defects appear not to be important.  

\begin{figure}[p] 
\begin{picture}(420,270)(0,0)
\put(-50,-50) {\includegraphics{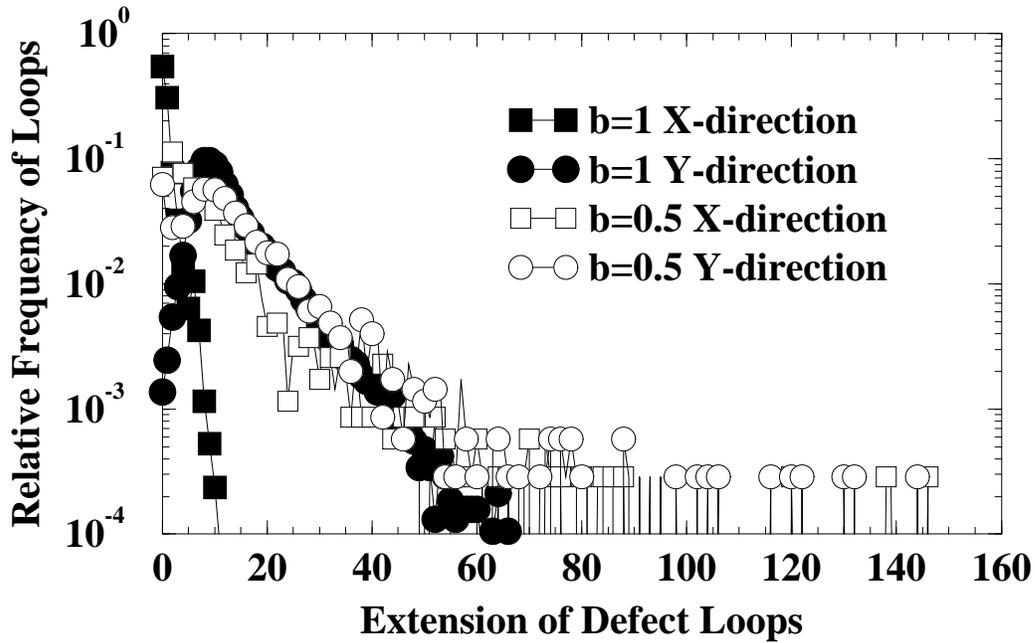}}
\end{picture}
\caption{Relative frequency of defect loops in space-time with extension $d_x$ in the
$x$-direction (squares) and $d_y$ in the $y$-direction (circles)
for $b=1$ (solid symbols) and $b=0.5$ (open symbols).
The other parameters are as in
fig.\protect{\ref{f:stabi}}.\protect{\label{f:size}}
}
\end{figure}

An interesting quantity is the `size' of the loops, i.e. 
the distance in the $X$- and in the $Y$-direction that is covered by a loop
during its life.  To determine this size it has to be taken into
account that, due to the periodic boundary conditions,
loops can `wrap around' the system and can, in principle, become larger than the
system size. The results are shown in fig.\ref{f:size}. Again, there
is a strong difference between the two parameter regimes. While for
$b=0.5$ there is a substantial number of loops that cover a large part of the system
in the $Y$- as well as the $X$-direction, 
the loops remain very restricted in the $X$-direction in the ordered regime, $b=1$. 
Thus, for $b=1$ the
defects  exhibit essentially only climb-motion while for $b=0.5$ they travel
in all directions. The latter changes the orientation of the
pattern and is responsible for the near-isotropy in the correlation 
function.

\section{Conclusion}

In this paper we have presented the results of numerical simulations of a model
for parametrically driven waves in a two-dimensional anisotropic system. 
Two types of behavior were found: in a regime in which in 1 dimension all phase
slips occur in sequential pairs the two-dimensional waves form an ordered stripe pattern
with a very slowly decaying correlation function; for smaller forcing, i.e. closer
to the parity-breaking instability, the waves form a very disordered pattern with
a rapidly decaying correlation function which is almost isotropic. An analysis of 
various statistics of the motion of the defects involved in the dynamics shows that
in the ordered regime they have a typical life-time whereas in the disordered regime
the distribution function is purely exponential. The number of defect pairs in 
the space-time loops traced out by
the defects increases strongly in the disordered regime. Another signature of the disordered
state is the extension of the loops. 
In the disordered regime they extend over large portions of the system in the
$X$- and the $Y$-direction, while in the ordered regime their extension in the $X$-direction
is strongly restricted. 

At the present it is not clear whether there is a true transition separating the 
two regimes or whether the change from one to the other is smooth. This needs to be
investigated. The extension in the $X$-direction should be a good diagnostic tool for this
question. 

An interesting question is whether there is any connection between
the disappearance of spatial order in this 
spatio-temporally chaotic system and two-dimensional melting \cite{Ne83}
and the Kosterlitz-Thouless transition \cite{KoTh73} in equilibrium systems.

HR gratefully acknowledges discussions with L. Kadanoff, R. Lipowsky and J. Marko. 
This work was supported by DOE through 
grant DE-FG02-92ER14303 and made use of the resources of
the Cornell Theory Center, which receives
major funding from NSF and New York State with additional support from the Advanced Research
Projects Agency, the National Center for Research Resources at the National Institutes
of Health, IBM Corporation and members of the Corporate Research Institute.


\bibliography{/home2/hermann/.bibfiles/journal}

\pagebreak

\newpage
\ \\
\newpage

\newpage
\ \\
\newpage

\newpage
\ \\
\newpage

\end{document}